\documentclass[runningheads]{llncs}
\usepackage[T1]{fontenc}
\usepackage{amsfonts}
\usepackage{makecell}
\usepackage{multirow}
\usepackage{graphicx}
\begin{document}
\sloppy
\title{Multi-Resolution Histopathology Patch Graphs for Ovarian Cancer Subtyping}
\titlerunning{Multi-Resolution Patch Graphs for Ovarian Cancer Subtyping}
%

\author{Jack Breen\inst{1}\orcidID{0000-0002-9020-3383} \and
Katie Allen\inst{2,3}\orcidID{0000-0002-5861-3745} \and
Kieran Zucker\inst{3}\orcidID{0000-0003-4385-3153} \and
Nicolas M. Orsi\inst{2,3}\orcidID{0000-0003-0890-0399} \and
Nishant Ravikumar\inst{1}\orcidID{0000-0003-0134-107X}}
\authorrunning{J. Breen et al.}
\institute{Centre for Computational Imaging and Simulation Technologies in Biomedicine (CISTIB), School of Computing, University of Leeds, UK\\
\email{\{scjjb,n.ravikumar\}@leeds.ac.uk} \and Leeds Institute of Medical Research at St James's, School of Medicine, \\University of Leeds, UK \and Leeds Cancer Centre, St James's University Hospital, Leeds, UK }

%
%

%
\maketitle              
\begin{abstract}

Computer vision models are increasingly capable of classifying ovarian epithelial cancer subtypes, but they differ from pathologists by processing small tissue patches at a single resolution. Multi-resolution graph models leverage the spatial relationships of patches at multiple magnifications, learning the context for each patch. In this study, we conduct the most thorough validation of a graph model for ovarian cancer subtyping to date. Seven models were tuned and trained using five-fold cross-validation on a set of 1864 whole slide images (WSIs) from 434 patients treated at Leeds Teaching Hospitals NHS Trust. The cross-validation models were ensembled and evaluated using a balanced hold-out test set of 100 WSIs from 30 patients, and an external validation set of 80 WSIs from 80 patients in the Transcanadian Study. The best-performing model, a graph model using 10x+20x magnification data, gave balanced accuracies of 73\%, 88\%, and 99\% in cross-validation, hold-out testing, and external validation, respectively. However, this only exceeded the performance of attention-based multiple instance learning in external validation, with a 93\% balanced accuracy. Graph models benefitted greatly from using the UNI foundation model rather than an ImageNet-pretrained ResNet50 for feature extraction, with this having a much greater effect on performance than changing the subsequent classification approach. The accuracy of the combined foundation model and multi-resolution graph network offers a step towards the clinical applicability of these models, with a new highest-reported performance for this task, though further validations are still required to ensure the robustness and usability of the models.

\end{abstract}
\section{Introduction}

Ovarian cancer is the eighth most common cancer in women worldwide \cite{Bray2024} and is characterised by heterogeneous histological subtypes. The five most common subtypes, which account for 90\% of all ovarian cancers, are high-grade serous (HGSC), low-grade serous (LGSC), clear cell (CCC), mucinous (MC), and endometrioid carcinomas (EC). These subtypes are distinct in their genetics, prognoses, and treatment options \cite{Kobel2008}, making their classification an essential component of ovarian cancer diagnosis. However, determining the subtype from standard histopathology samples can be a difficult task with a high level of inter-observer discordance \cite{Kobel2014}. 
Thus, it is common for pathologists to request ancillary tests and second opinions to ensure an accurate diagnosis, slowing the diagnostic pathway and increasing costs.    

Artificial intelligence (AI) models have been proposed as potential assistive tools for pathological diagnosis. While some tools are starting to receive regulatory approval \cite{Matthews2024} and be clinically validated \cite{Raciti2023}, this is not the case for ovarian cancer subtyping models. Previous research in this area has involved model prototyping with relatively small homogeneous datasets \cite{Breen2023review}, though some recent studies have included larger and more diverse datasets \cite{Asadi2024,Breen2024foundation,Farahani2022}. The highest reported performances for five-class classification were 81\% and 93\% balanced accuracies \cite{Farahani2022,Breen2024foundation}, compared to median pathologist concordance rates of 78-86\% from histopathology alone, and 90\% with the addition of immunohistochemical information \cite{Kobel2014}. 

Ovarian cancer resection whole slide images (WSIs) contain billions of pixels and are stored in files that are typically 1-4GB in size. Traditional computer vision models cannot handle such large images, so information is typically extracted from small patches and aggregated to the WSI level in a process called `multiple instance learning' (MIL). 
These models often treat patches as being functionally independent of one another, such as in attention-based multiple instance learning (ABMIL) \cite{Ilse2018}, where a weighted average of patch embeddings is taken to form a WSI embedding. 

It can instead be beneficial to leverage the spatial or semantic relationships between patches by integrating graphs \cite{Scarselli2008} or transformers \cite{Vaswani2017} into MIL models \cite{Bazargani2024,Chen2021,Guo2023,Shao2021,Tu2019}. Only one previous study has applied such a method to ovarian cancer subtyping \cite{Mirabadi2024}, where it was reported that a novel multi-resolution graph network gave a better balanced accuracy than previous methods, including ABMIL, TransMIL \cite{Shao2021}, and single-magnification graph models. However, this study used only a single set of model hyperparameters for all models, and a single dataset for evaluation, making it unclear whether the models were properly tuned to the given task and data, and unclear how well the models would generalise.   

In this paper, we present the most thorough evaluation of a graph model for ovarian cancer subtyping to date, including hyperparameter tuning and both hold-out and external validations. To the best of our knowledge, it is also the first graph-based MIL model to be conducted using features from the vision transformer (ViT)-based histopathology foundation model, UNI \cite{Chen2024}.

\section{Methods}
\subsection{Data}

The training dataset comprised 1864 ovarian carcinoma resection WSIs from 434 patients at Leeds Teaching Hospitals NHS Trust. These were retrospectively collected by a pathologist, who confirmed the original diagnoses made by a gynaecological subspecialty expert. A class-balanced independent hold-out test dataset was collected following the same procedure as in the training set, comprising 100 WSIs from 30 patients. All WSIs were generated from slides of haematoxylin and eosin (H\&E)-stained formalin-fixed and paraffin-embedded (FFPE) adnexal tissue which was digitised at 40x magnification on an Aperio AT2 scanner. The training set slides contained both primary surgery samples and interval debulking surgery samples, while the hold-out test set contained primary samples alone. An external dataset of 80 WSIs from 80 ovarian cancer patients was sourced from the Transcanadian Study \cite{Kobel2010}, with these provided at 20x magnification.

\begin{table}[htbp]
\begin{center}
\caption{Class breakdown of the datasets used in model training and validation, with the number of whole slide images (WSIs) and the number of patients indicated.}
\label{table:dataset}
\begin{tabular}{c|ccc}

\textbf{ \makecell{Ovarian Cancer \\ Histological Subtype} \hspace{0.1cm}}                     & \textbf{\hspace{0.1cm} \makecell{Training Set} \hspace{0.1cm}}       & \textbf{\hspace{0.1cm} \makecell{Independent \\ Test Set} \hspace{0.1cm}} & \textbf{\hspace{0.1cm} \makecell{External \cite{Kobel2010} \\ Test Set} }      \\ \hline \hline 
\makecell{High-Grade Serous \\ Carcinoma (HGSC)}    & \makecell{1266 WSIs, \\ 308 patients}                      & \makecell{20 WSIs, \\ 7 patients}   & \makecell{30 WSIs, \\ 30 patients}                      \\ \hline
\makecell{Endometrioid \\ Carcinoma (EC)}           & \makecell{209 WSIs, \\ 38 patients}                       & \makecell{20 WSIs, \\ 5 patients}   & \makecell{11 WSIs, \\ 11 patients}                     \\ \hline
\makecell{Clear Cell \\ Carcinoma (CCC)}            & \makecell{198 WSIs, \\ 45 patients}                       & \makecell{20 WSIs, \\ 7 patients}    & \makecell{20 WSIs, \\ 20 patients}                    \\ \hline
\makecell{Low-Grade Serous \\ Carcinoma (LGSC)}     & \makecell{92 WSIs, \\ 21 patients}                         & \makecell{20 WSIs, \\ 6 patients}    & \makecell{9 WSIs, \\ 9 patients}                    \\ \hline
\makecell{Mucinous \\ Carcinoma (MC)}               & \makecell{99 WSIs, \\ 22 patients}                        & \makecell{20 WSIs, \\ 5 patients}   & \makecell{10 WSIs, \\ 10 patients}                     \\ \hline \hline
\textbf{Overall} & \textbf{\makecell{1864 WSIs, \\ 434 patients}} & \textbf{\makecell{100 WSIs, \\ 30 patients}} & \textbf{\makecell{80 WSIs, \\ 80 patients}} \\ \hline 
\end{tabular}%
\end{center}
\end{table}

\subsection{Modelling}

\begin{figure}[htb]
\centering
\includegraphics[width=0.8\textwidth]{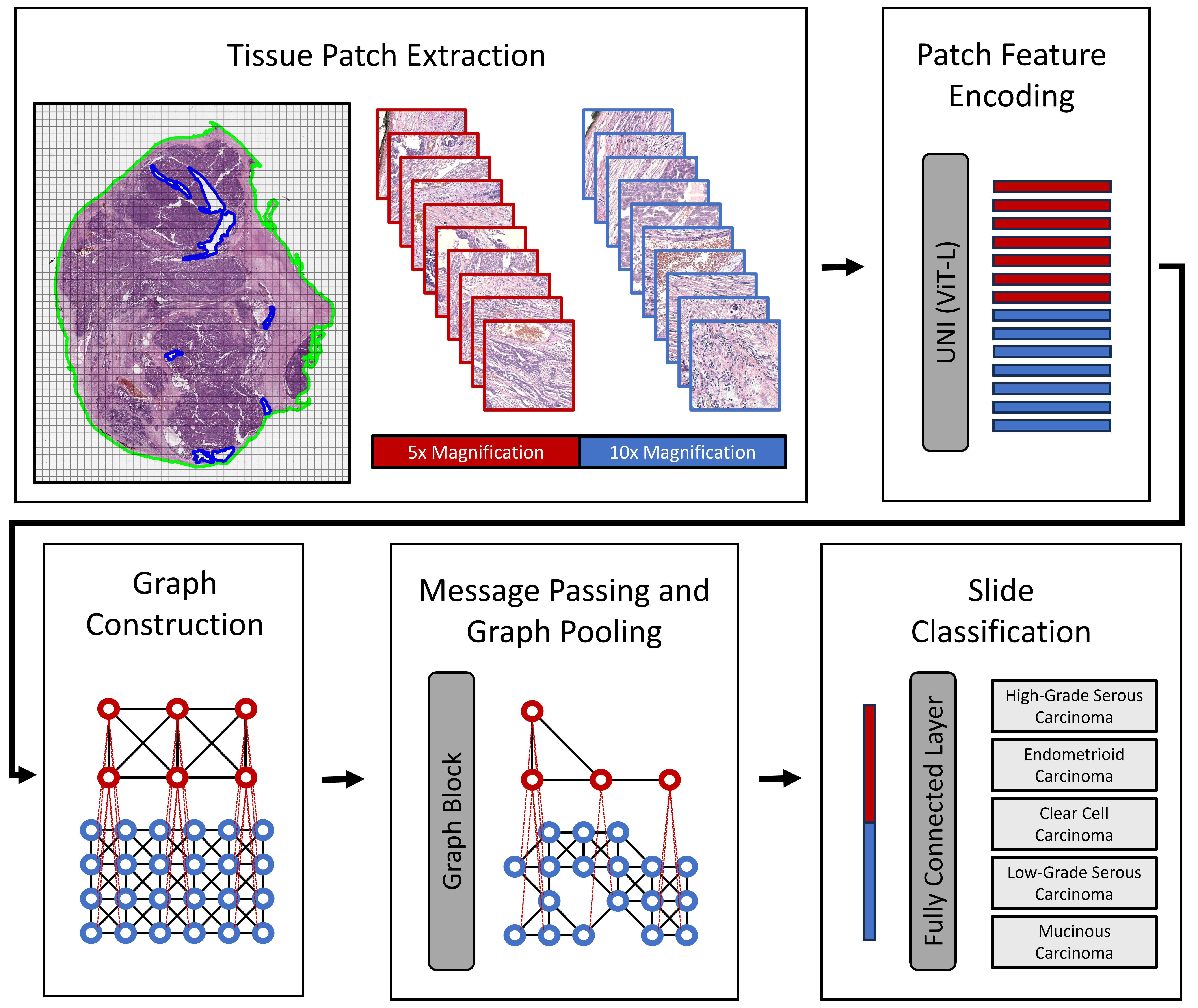}
\caption{Multi-resolution graph network model pipeline for whole slide image classification, illustrated using 5x and 10x magnification tissue patches. Graph blocks were composed of at least one GATv2 message-passing layer \cite{Brody2021} followed by a SAGPool graph pooling layer \cite{Lee2019}.} \label{fig:pipeline}
\end{figure}

A multi-stage WSI classification pipeline was employed (shown in Figure \ref{fig:pipeline}). First, the tissue region was segmented from plain background and tissue patches were selected. Patch-level features were then extracted using a pre-trained model. Graphs were constructed based on the spatial arrangement of the patches, with message-passing layers used to share information between neighbouring patches. Graphs were pooled (often multiple times) to reduce their size, and the remaining features were averaged to generate a WSI-level encoding for five-class classification through a fully connected neural network. 

Tissue segmentation was performed using a basic saturation thresholding approach (from CLAM \cite{Lu2021}). Non-overlapping patches were extracted to cover the entire tissue area, with the patch size determined to give 256x256 pixels after downsampling from the 40x native magnification to the chosen apparent magnification. This required taking 512x512 patches for 20x, 1024x1024 patches for 10x, and 2048x2048 pixel patches for 5x, with each doubling of the apparent magnification quadrupling the number of resulting patches. This was the only step that differed for the external dataset, with smaller patches required before downsampling given the lower original scanning magnification. Features were extracted from all downsampled 256x256 pixel patches using the histopathology-pretrained vision transformer `foundation model', UNI \cite{Chen2024}, which has been demonstrated to give state-of-the-art performance for ovarian cancer subtyping with an ABMIL classifier \cite{Breen2024foundation}. For comparison, the features were also extracted using an ImageNet-pretrained ResNet50 convolutional neural network encoder \cite{He2016,Russakovsky2015}.        

Graphs were constructed such that each patch was connected to any other patch within a given radius, which was set to allow connections to first-order lateral and diagonal neighbours. Connections were also made between patches showing the same tissue at different magnifications, with each low-magnification patch connected to four high-magnification patches. The graph model consisted of graph blocks, which were composed of at least one graph attention (GATv2) convolution layer \cite{Brody2021} for message passing, followed by a ReLU activation, before a final self-attention graph pooling (SAGPool) \cite{Lee2019} layer to reduce the number of nodes in the graph. The outputs of each graph block were pooled using both mean and max pooling across all remaining nodes, and these pooled feature sets were concatenated together to make a double-length feature set. These graph block outputs were summed to form a WSI-level feature set (to which dropout was applied during training), and finally, these were classified through a single fully connected network layer with five output neurons corresponding to the five ovarian cancer subtypes.     


One complexity in extending graph networks to multiple resolutions is in handling the features at multiple magnifications. Different biological entities are represented by features at different magnifications; thus, it may be naive to have a shared latent space across magnifications. Further, given the greater quantity of patches at higher magnifications, these may have an undue influence in a shared feature space. Previous studies have concatenated features from different resolutions \cite{Bazargani2024,Mirabadi2024}, though it is unclear whether this is the best strategy. We compared the `naive' approach (in which features are not separated by magnification) to two concatenation approaches which differ in their initialisation. Each graph node initially represented a tissue patch at a single magnification, so the part of the embedding representing the other magnification was either initialised as a zero vector (`concat\_zero') or as the average of all patch features at the given magnification (`concat\_avg'). This issue was avoided in the previous ovarian cancer graph model \cite{Mirabadi2024} by analysing only one high-resolution patch within each lower-resolution patch, though this discarded most of the high-resolution tissue, whereas the currently presented model used all available tissue at each magnification.


\subsection{Training and Validation Procedures}

Graph model hyperparameters were tuned through an iterative grid search procedure, with up to two hyperparameters adjusted at a time and all others frozen at their previous best, starting from the optimal hyperparameters of a previous study \cite{Breen2024foundation} in which ABMIL was tuned for the same task and dataset as used in this study. Each hyperparameter configuration was evaluated using the average balanced cross-entropy loss across the validation sets of five-fold cross-validation. Models were trained using an Adam optimizer, with class-balanced sampling used to account for the imbalance in the training set. At least 100 unique configurations were evaluated during the tuning of each graph-based model, with the ABMIL hyperparameters instead taken from the previous study \cite{Breen2024foundation}. Hyperparameter tuning is described further in Supplementary Section \ref{app:hyperparams}.

Seven total models were evaluated to compare different feature extractors, magnifications, and classifiers. The baseline multi-resolution graph model combined 5x and 10x magnifications, which have each previously been found to give an accurate and efficient classification of ovarian cancer subtypes \cite{Breen2024ISBI}. Comparisons to different models and magnifications were conducted using a multi-resolution graph at 10x and 20x magnifications, a single-magnification graph at 10x, and ABMIL at 10x. In another comparison, the UNI vision transformer foundation model features used in the baseline were changed to features from an ImageNet-pretrained ResNet50. Finally, to compare different multi-resolution feature spaces, the baseline approach of using separate magnification-specific features with average initialisation (concat\_avg) was compared to the separate features with zero initialisation (concat\_zero) and to a magnification-agnostic combined feature space (naive). Paired t-tests were used to statistically compare each model to the baseline model across the five cross-validation folds, with p-values adjusted for multiple testing \cite{Benjamini1995}. 

\section{Results}

\begin{figure}[h]
\includegraphics[width=\textwidth]{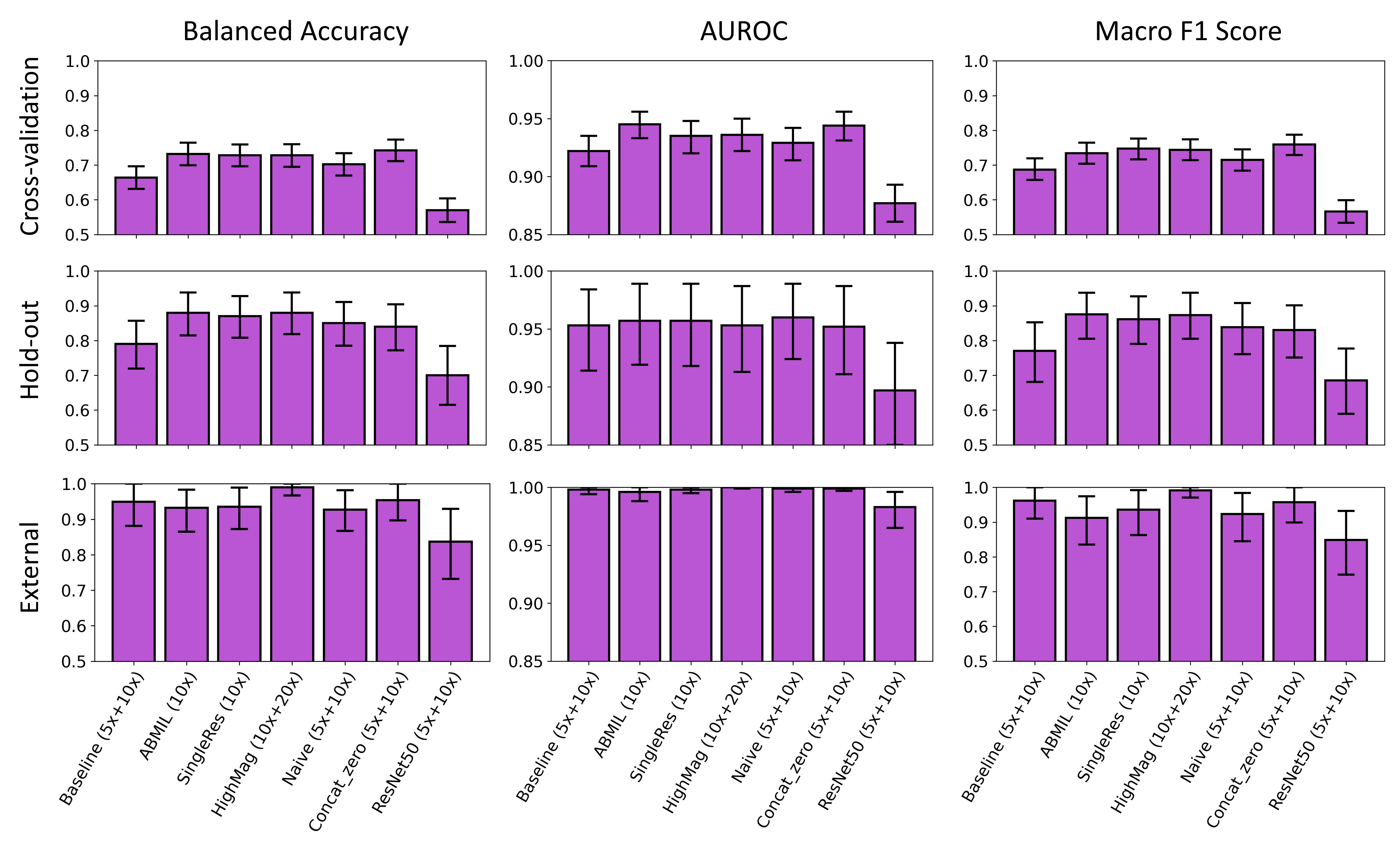}
\caption{Results of all seven ovarian cancer subtyping models in cross-validation, internal hold-out testing, and external validation on the Transcandian Study dataset \cite{Kobel2010}. In hold-out testing and external validation, predictions were made using an average ensemble of the five cross-validation model predictions. Results are shown as the mean and 95\% confidence interval from 10,000 iterations of bootstrapping.} \label{fig:results}
\end{figure}

All results are shown in Figure \ref{fig:results}, with full breakdowns and p-values in Supplementary Section \ref{app:results}. The best-performing model in cross-validation was the multi-resolution graph with a zero-initialised feature space, with a balanced accuracy of 74.2\%, AUROC of 0.944 (second-best, behind the 0.945 of ABMIL), and F1 score of 0.759. In hold-out testing, ABMIL performed best, with a balanced accuracy of 88.0\%, AUROC of 0.957 (second-best, behind the 0.960 of the naive graph model), and an F1 score of 0.875. In external validation, the 10x+20x magnification graph performed best by all metrics, with a balanced accuracy of 99.0\%, AUROC of 1.000, and an F1 score of 0.991.

No single model gave the best performance for all metrics across all evaluations, though the 10x+20x magnification model was the most consistent, never more than 0.014 behind the best for any given metric. In cross-validation and hold-out testing, this model was slightly outperformed by ABMIL and the graph model with the naive feature space, but in external validation the 10x+20x graph model was best by a clear margin (3.6\% balanced accuracy, 0.001 AUROC, 0.029 F1 score). The model using the ImageNet-pretrained ResNet50 encoder performed worst in every evaluation, indicating the clear benefit of the newer foundation model features. In fact, the selection of an appropriate feature encoder had a much greater effect on model performance than the subsequent MIL modelling approach, with ABMIL being competitive with the best graph models by most metrics.


The effects of modelling with different magnifications were not consistent across all evaluations. The 10x graph model had significantly higher balanced accuracy and F1 scores than the 5x+10x baseline in cross-validation and hold-out testing but performed slightly worse in external validation. The 10x+20x model outperformed the 5x+10x model in all evaluations, though this difference was only statistically significant for the hold-out AUROC. The 10x+20x model gave similar performance to the 10x-only model in cross-validation and hold-out testing but performed much better in external validation. 

It was unclear which multi-resolution feature space was best overall, with each of the three approaches best in some cases. The baseline average-initialised feature space was generally the worst of the three, with the naive combined feature space best in hold-out testing, and the zero-initialised features best in cross-validation and external validation. As such, the optimal 10x+20x model may have benefitted further from initialisation using zeroes rather than averages.  



\section{Discussion}

Overall, the results indicate that multi-resolution graph models can offer improvements to ovarian carcinoma subtyping. In particular, the 10x+20x magnification model achieved near-perfect performance in external validation, making this the greatest reported performance for this task to date \cite{Breen2023review}. However, multi-resolution graph models did not offer a clear benefit over ABMIL in internal validations and, considering the relatively small size of the external validation set, it is unclear how great a benefit these models offer overall. 


As in the previous study using these datasets \cite{Breen2024foundation}, performance was greater in hold-out testing than cross-validation, and greater still in external testing. While it is not exactly clear why this is the case, several factors may have had an impact. The hold-out and external validations used model ensembles and did not include interval debulking surgery samples that can be more diagnostically challenging. The external validation data also appeared to have given a “best-case scenario”, with a single slide per patient and a high proportion of carcinoma within the tissue present on the high-quality WSI. Further, the different scanning magnifications may have led to a difference in image quality after downsampling, though such an effect was not visibly apparent.  


In external validation, all models achieved AUROC scores between 0.983 and 1.000, despite the balanced accuracy and F1 scores varying from 83.7\%-99.0\% and 0.849-0.991, respectively. 
There were also several instances in internal testing where models had highly similar AUROC scores but clearly distinct scores by the other metrics.  
This highlights the limitation of the AUROC for imbalanced multi-class classification, given its reporting similarly high performances for all models despite obvious differences in the balanced accuracy and F1 scores, which are more representative of clinical utility. 

The five-class balanced accuracies of 88\% and 99\% in hold-out and external validations may be sufficient for clinical assistance tools, but some limitations remain. These validations used data from only 30 and 80 patients, respectively, so cannot represent the vast array of variability seen in clinical diagnostic cases. The models are also currently incapable of indicating uncertainty, providing thorough explanations of classification decisions, or coping with tissue which does not contain one of the five most common subtypes of ovarian carcinoma (e.g. non-malignant tissue, carcinosarcomas and non-epithelial malignancies). The large vision transformer feature extractors and multi-resolution graph networks also carry a heavy computational burden, which is likely to lead to logistical difficulties in deploying such models in the clinical setting. None of these issues are insurmountable, and when they are overcome, these models could be invaluable as diagnostic assistive tools offering a rapid second opinion to pathologists.       


\section{Conclusion}
Overall, we have shown that multi-resolution graph models can improve the accuracy of ovarian carcinoma subtyping at the whole-slide level above the previous state-of-the-art, though it was not beneficial in all validations. In an external validation of 80 WSIs, a graph model achieved a near-perfect 99\% balanced-accuracy, but in internal testing this was only 88\%, which was no greater than ABMIL. The best model combined data at 10x and 20x magnifications, which was better than either using lower magnifications or only using 10x magnification data. While the benefit provided by graph models may offer a clinically useful second opinion to pathologists, more extensive validations are required to understand the reasons underlying performance variability across different datasets and to improve model consistency.   



\begin{credits}
\subsubsection{\ackname} JB is supported by the UKRI Engineering and Physical Sciences Research Council (EPSRC) [EP/S024336/1]. KA is supported by the Tony Bramall Charitable Trust. The funders had no role in influencing the content of this research. 
This study was conducted retrospectively using human subject data and received approval from the Wales Research Ethics Committee [18/WA/0222] and the Confidentiality Advisory Group [18/CAG/0124]. 
External data were downloaded from \url{https://www.medicalimageanalysis.com/data/ovarian-carcinomas-histopathology-dataset} (last accessed 09/04/24). 
All code used in this research is available at
\url{https://github.com/scjjb/MultiscalePathGraph}. 
For the purpose of open access, the author has applied a Creative Commons Attribution (CC BY) licence to any Author Accepted Manuscript version arising from this submission. Experiments were conducted using an NVIDIA A100 GPU and 32 AMD EPYC7742 CPUs @3.4GHz.

\subsubsection{\discintname}
NMO's fellowship is funded by 4D Path. All other authors declare no conflicts of interest.

\end{credits}
%
%
%
%
\bibliographystyle{splncs04}
\bibliography{library}

\begin{thebibliography}{10}
\providecommand{\url}[1]{\texttt{#1}}
\providecommand{\urlprefix}{URL }
\providecommand{\doi}[1]{https://doi.org/#1}

\bibitem{Asadi2024}
Asadi-Aghbolaghi, M., Farahani, H., Zhang, A., Akbari, A., Kim, S., Chow, A., Dane, S., Consortium, O.C., Consortium, O., G~Huntsman, D., et~al.: Machine learning-driven histotype diagnosis of ovarian carcinoma: Insights from the ocean ai challenge. medRxiv pp. 2024--04 (2024)

\bibitem{Bazargani2024}
Bazargani, R., Fazli, L., Gleave, M., Goldenberg, L., Bashashati, A., Salcudean, S.: Multi-scale relational graph convolutional network for multiple instance learning in histopathology images. Medical Image Analysis  \textbf{96},  103197 (2024)

\bibitem{Benjamini1995}
Benjamini, Y., Hochberg, Y.: Controlling the false discovery rate: a practical and powerful approach to multiple testing. Journal of the Royal statistical society: series B (Methodological)  \textbf{57}(1),  289--300 (1995)

\bibitem{Bray2024}
Bray, F., Laversanne, M., Sung, H., Ferlay, J., Siegel, R.L., Soerjomataram, I., Jemal, A.: Global cancer statistics 2022: Globocan estimates of incidence and mortality worldwide for 36 cancers in 185 countries. CA: A Cancer Journal for Clinicians pp. 1--35 (2024). \doi{https://doi.org/10.3322/caac.21834}, \url{https://acsjournals.onlinelibrary.wiley.com/doi/abs/10.3322/caac.21834}

\bibitem{Breen2023review}
Breen, J., Allen, K., Zucker, K., Adusumilli, P., Scarsbrook, A., Hall, G., Orsi, N.M., Ravikumar, N.: Artificial intelligence in ovarian cancer histopathology: a systematic review. NPJ Precision Oncology  \textbf{7}(1), ~83 (2023)

\bibitem{Breen2024foundation}
Breen, J., Allen, K., Zucker, K., Godson, L., Orsi, N.M., Ravikumar, N.: Histopathology foundation models enable accurate ovarian cancer subtype classification. arXiv preprint arXiv:2405.09990  (2024)

\bibitem{Breen2024ISBI}
Breen, J., Allen, K., Zucker, K., Orsi, N.M., Ravikumar, N.: Reducing histopathology slide magnification improves the accuracy and speed of ovarian cancer subtyping. arXiv preprint arXiv:2311.13956  (2023)

\bibitem{Brody2021}
Brody, S., Alon, U., Yahav, E.: How attentive are graph attention networks? arXiv preprint arXiv:2105.14491  (2021)

\bibitem{Chen2024}
Chen, R.J., Ding, T., Lu, M.Y., Williamson, D.F., Jaume, G., Song, A.H., Chen, B., Zhang, A., Shao, D., Shaban, M., et~al.: Towards a general-purpose foundation model for computational pathology. Nature Medicine pp. 1--13 (2024)

\bibitem{Chen2021}
Chen, R.J., Lu, M.Y., Shaban, M., Chen, C., Chen, T.Y., Williamson, D.F., Mahmood, F.: Whole slide images are 2d point clouds: Context-aware survival prediction using patch-based graph convolutional networks. In: Medical Image Computing and Computer Assisted Intervention--MICCAI 2021: 24th International Conference, Strasbourg, France, September 27--October 1, 2021, Proceedings, Part VIII 24. pp. 339--349. Springer (2021)

\bibitem{Farahani2022}
Farahani, H., Boschman, J., Farnell, D., Darbandsari, A., Zhang, A., Ahmadvand, P., Jones, S.J., Huntsman, D., K{\"o}bel, M., Gilks, C.B., et~al.: Deep learning-based histotype diagnosis of ovarian carcinoma whole-slide pathology images. Modern Pathology  \textbf{35}(12),  1983--1990 (2022)

\bibitem{Guo2023}
Guo, Z., Zhao, W., Wang, S., Yu, L.: Higt: Hierarchical interaction graph-transformer for whole slide image analysis. In: International Conference on Medical Image Computing and Computer-Assisted Intervention. pp. 755--764. Springer (2023)

\bibitem{He2016}
He, K., Zhang, X., Ren, S., Sun, J.: Deep residual learning for image recognition. In: Proceedings of the IEEE conference on computer vision and pattern recognition. pp. 770--778 (2016)

\bibitem{Ilse2018}
Ilse, M., Tomczak, J., Welling, M.: Attention-based deep multiple instance learning. In: International conference on machine learning. pp. 2127--2136. PMLR (2018)

\bibitem{Kobel2014}
K{\"o}bel, M., Bak, J., Bertelsen, B.I., Carpen, O., Grove, A., Hansen, E.S., Levin~Jakobsen, A.M., Lidang, M., M{\aa}sb{\"a}ck, A., Tolf, A., et~al.: Ovarian carcinoma histotype determination is highly reproducible, and is improved through the use of immunohistochemistry. Histopathology  \textbf{64}(7),  1004--1013 (2014)

\bibitem{Kobel2010}
K{\"o}bel, M., Kalloger, S.E., Baker, P.M., Ewanowich, C.A., Arseneau, J., Zherebitskiy, V., Abdulkarim, S., Leung, S., Duggan, M.A., Fontaine, D., et~al.: Diagnosis of ovarian carcinoma cell type is highly reproducible: a transcanadian study. The American journal of surgical pathology  \textbf{34}(7),  984--993 (2010)

\bibitem{Kobel2008}
K{\"o}bel, M., Kalloger, S.E., Boyd, N., McKinney, S., Mehl, E., Palmer, C., Leung, S., Bowen, N.J., Ionescu, D.N., Rajput, A., et~al.: Ovarian carcinoma subtypes are different diseases: implications for biomarker studies. PLoS medicine  \textbf{5}(12), ~e232 (2008)

\bibitem{Lee2019}
Lee, J., Lee, I., Kang, J.: Self-attention graph pooling. In: International conference on machine learning. pp. 3734--3743. PMLR (2019)

\bibitem{Lu2021}
Lu, M.Y., Williamson, D.F., Chen, T.Y., Chen, R.J., Barbieri, M., Mahmood, F.: Data-efficient and weakly supervised computational pathology on whole-slide images. Nature biomedical engineering  \textbf{5}(6),  555--570 (2021)

\bibitem{Matthews2024}
Matthews, G.A., McGenity, C., Bansal, D., Treanor, D.: Public evidence on ai products for digital pathology. medRxiv pp. 2024--02 (2024)

\bibitem{Mirabadi2024}
Mirabadi, A.K., Archibald, G., Darbandsari, A., Contreras-Sanz, A., Nakhli, R.E., Asadi, M., Zhang, A., Gilks, C.B., Black, P., Wang, G., et~al.: {GRASP}: Graph-structured pyramidal whole slide image representation. arXiv preprint arXiv:2402.03592  (2024)

\bibitem{Raciti2023}
Raciti, P., Sue, J., Retamero, J.A., Ceballos, R., Godrich, R., Kunz, J.D., Casson, A., Thiagarajan, D., Ebrahimzadeh, Z., Viret, J., et~al.: Clinical validation of artificial intelligence--augmented pathology diagnosis demonstrates significant gains in diagnostic accuracy in prostate cancer detection. Archives of Pathology \& Laboratory Medicine  \textbf{147}(10),  1178--1185 (2023)

\bibitem{Russakovsky2015}
Russakovsky, O., Deng, J., Su, H., Krause, J., Satheesh, S., Ma, S., Huang, Z., Karpathy, A., Khosla, A., Bernstein, M., et~al.: Imagenet large scale visual recognition challenge. International journal of computer vision  \textbf{115},  211--252 (2015)

\bibitem{Scarselli2008}
Scarselli, F., Gori, M., Tsoi, A.C., Hagenbuchner, M., Monfardini, G.: The graph neural network model. IEEE transactions on neural networks  \textbf{20}(1),  61--80 (2008)

\bibitem{Shao2021}
Shao, Z., Bian, H., Chen, Y., Wang, Y., Zhang, J., Ji, X., et~al.: Transmil: Transformer based correlated multiple instance learning for whole slide image classification. Advances in neural information processing systems  \textbf{34},  2136--2147 (2021)

\bibitem{Tu2019}
Tu, M., Huang, J., He, X., Zhou, B.: Multiple instance learning with graph neural networks. arXiv preprint arXiv:1906.04881  (2019)

\bibitem{Vaswani2017}
Vaswani, A., Shazeer, N., Parmar, N., Uszkoreit, J., Jones, L., Gomez, A.N., Kaiser, {\L}., Polosukhin, I.: Attention is all you need. Advances in neural information processing systems  \textbf{30} (2017)

\end{thebibliography}

\newpage
\appendix

\section{Supplementary Hyperparameter Tuning}\label{app:hyperparams}

A total of 13 hyperparameters were tuned for the graph models, which are grouped into those influencing the learning rate, the optimizer, the regularisation, and the model architecture. The learning rate hyperparameters set the initial value, the decay multiplier, and the decay patience (in epochs). The optimizer hyperparameters, $\beta_1$, $\beta_2$, and $\epsilon$, controlled the decay of the first and second moments, and the optimization stability, respectively. Three types of regularisation were used during model training, which were weight decay, parameter dropout, and patch dropout (defined by the maximum number of patches randomly selected per slide). Hyperparameters relating to the model architecture controlled the number of message-passing layers per graph block, the number of graph blocks (and hence the number of pooling layers), the graph pooling factor, and the embedding size per magnification. 

The best hyperparameters from tuning each model are shown in Table \ref{tab:hyperparams}. The smallest tuned classifiers were the single-resolution (10x graph, 0.5M; 10x ABMIL, 0.8M) and combined feature space methods (naive features, 0.7M), followed closely by the zero-initialised model (1.2M) and the 10x+20x graph (1.2M), with the largest being the baseline model with average initialisation (7.9M) and the ResNet-based classifier (10.5M). In most cases, the classifier was smaller than the respective feature extractor, with the UNI model having 303M parameters and the ResNet50 having 9M.


\begin{table}
    \centering
    \caption{Optimal hyperparameters for each model found through an iterative grid search on the validation sets from five-fold cross-validation. These are grouped into hyperparameters relating to the learning rate, Adam optimizer, regularisation, and model architecture.}
    \label{tab:hyperparams}
    \resizebox{\columnwidth}{!}{%
    \begin{tabular}{|c|c|c|c|c|c|c|c|} \hline 
         \textbf{ \hspace{0.05cm}Hyperparameter \hspace{0.05cm}} & \textbf{\makecell{ \hspace{0.05cm} Baseline \hspace{0.05cm} \\ graph }}& \textbf{\makecell{ABMIL \\ \hspace{0.05cm} 10x only \hspace{0.05cm}}}& \textbf{\makecell{Graph \\ \hspace{0.05cm} 10x only \hspace{0.05cm} }}& \textbf{\makecell{Graph \\ \hspace{0.05cm} 10x+20x \hspace{0.05cm}}}& \textbf{\makecell{Naive \\ \hspace{0.05cm} features \hspace{0.05cm}}}& \textbf{\makecell{ \hspace{0.025cm} Concat\_zero \hspace{0.025cm} \\ features}}& \textbf{\makecell{ \hspace{0.05cm} ImageNet- \hspace{0.05cm}\\ ResNet50}}\\ \hline \hline 
         Learning rate (LR)&  1e-4&  1e-5&  5e-5&  1e-4&  2e-4&  1e-4& 2e-3\\ \hline 
         LR decay&  0.9&  0.75&  0.9&  0.9&  0.45&  0.9&0.6\\\hline
         LR decay patience&  10&  10&  10&  20&  15&  15&20\\\hline \hline
         Beta1&  0.9&  0.9&  0.95&  0.95&  0.9&  0.95& 0.8\\ \hline 
         Beta2&  0.9999&  0.999&  0.999&  0.999&  0.99999&  0.99& 0.95\\ \hline 
         Epsilon&  1e-5&  1e-5&  1e-7&  1e-7&  1e-7&  1e-7& 1e-2\\ \hline \hline
         Dropout&  0.2&  0.0&  0.0&  0.1&  0.2&  0.4& 0.2\\ \hline 
         Weight decay&  1e-2&  1e-3&  1e-1&  1e-2&  1e-3&  1e-2& 1e-3\\ \hline 
         Max patches&  6000&  1000&  4000&  14000&  4000&  5000& 5000\\ \hline \hline
         Message passings&  3&  NA&  1&  1&  1&  1& 1\\ \hline 
         Graph poolings&  4&  NA&  1&  2&  2&  2& 4\\ \hline
         Pooling factor&  0.9&  NA&  0.6&  0.6&  0.45&  0.75&0.6\\\hline
         Embedding size&  512&  512&  256&  256&  256&  256&1024\\\hline
    \end{tabular}
    }
\end{table}

\newpage
\section{Supplementary Results Tables} \label{app:results}
\begin{table}[h]
    \centering
    \caption{Cross-validation results shown as the mean and 95\% confidence intervals generated by 10,000 iterations of bootstrapping. The best results are indicated in \textbf{bold}.}
    \label{tab:results-cv}
    \resizebox{\columnwidth}{!}{%
    \begin{tabular}{|c|c|c|c|} \hline 
           Model&  Balanced Accuracy&  AUROC& F1 Score\\ \hline \hline
 Baseline graph 5x + 10x& 66.4\% (63.1-69.6\%) & 0.922 (0.909-0.935) & 0.687 (0.657-0.719) \\ \hline \hline
 ABMIL 10x only& 73.2\% (69.9-76.4\%)& \textbf{0.945} (0.933-0.956) & 0.734 (0.704-0.764)\\ \hline 
           Graph 10x only& 72.8\% (69.6-75.9\%) & 0.935 (0.920-0.948) & 0.747 (0.716-0.776) \\ \hline
           Graph 10x + 20x& 72.8\% (69.5-76.0\%) & 0.936 (0.922-0.950) & 0.744 (0.714-0.774) \\ \hline \hline
           Naive feature space& 70.2\% (67.0-73.4\%) & 0.929 (0.914-0.942) & 0.715 (0.684-0.745) \\ \hline 
           Concat\_zero feature space& \textbf{74.2\%} (71.1-77.3\%) & 0.944 (0.931-0.956) & \textbf{0.759} (0.729-0.787) \\ \hline \hline
           ImageNet-ResNet50 features& 57.0\% (53.6-60.4\%) & 
0.877 (0.861-0.893) & 0.566 (0.534-0.599) \\ \hline 
           
    \end{tabular}
    }
\end{table}

\begin{table}[h]
    \centering
    \caption{Hold-out testing results (ensembled across the cross-validation folds) shown as the mean and 95\% confidence intervals generated by 10,000 iterations of bootstrapping. The best results are indicated in \textbf{bold}.}
    \label{tab:results-holdout}
    \resizebox{\columnwidth}{!}{%
    \begin{tabular}{|c|c|c|c|} \hline 
           Model&  Balanced Accuracy&  AUROC& F1 Score\\ \hline \hline
 Baseline graph 5x + 10x& 79.0\% (71.9-85.7\%) & 0.953 (0.914-0.984) & 0.770 (0.681-0.852) \\ \hline \hline
 ABMIL 10x only& \textbf{88.0\%} (81.5-93.8\%) & 0.957 (0.919-0.989) & \textbf{0.875} (0.805-0.937) \\ \hline 
           Graph 10x only& 87.0\% (80.8-92.8\%) & 
0.957 (0.918-0.989) & 0.861 (0.790-0.927) \\ \hline
           Graph 10x + 20x& \textbf{88.0\%} (81.8-93.8\%) & 0.953 (0.913-0.987) & 0.873 (0.805-0.937) \\ \hline \hline
           Naive feature space& 85.0\% (78.5-91.1\%) & \textbf{0.960} (0.924-0.989) & 0.838 (0.761-0.908) \\ \hline 
           Concat\_zero feature space& 84.0\% (77.2-90.4\%) & 0.952 (0.911-0.987) & 0.830 (0.751-0.901) \\ \hline \hline
           ImageNet-ResNet50 features& 70.0\% (61.5-78.4\%) & 0.897 (0.850-0.938) & 0.685 (0.589-0.777) \\ \hline 
           
    \end{tabular}
    }
\end{table}

\begin{table}
    \centering
    \caption{External validation results (ensembled across the cross-validation folds) shown as the mean and 95\% confidence intervals generated by 10,000 iterations of bootstrapping. The best results are indicated in \textbf{bold}.}
    \label{tab:results-external}
    \resizebox{\columnwidth}{!}{%
    \begin{tabular}{|c|c|c|c|} \hline 
           Model&  Balanced Accuracy&  AUROC& F1 Score\\ \hline \hline
 Baseline graph 5x + 10x& 94.9\% (88.1-100.0\%) & 0.998 (0.994-1.000) & 0.962 (0.910-1.000) \\ \hline \hline
 ABMIL 10x only& 93.2\% (86.5-98.3\%) & 0.996 (0.988-1.000) & 0.912 (0.835-0.974) \\ \hline 
           Graph 10x only& 93.5\% (87.2-98.9\%) & 0.998 (0.995-1.000) & 0.936 (0.863-0.992) \\ \hline
           Graph 10x + 20x& \textbf{99.0\%} (96.7-100.0\%) & \textbf{1.000} (0.999-1.000) & \textbf{0.991} (0.971-1.000) \\ \hline \hline
           Naive feature space& 92.7\% (86.7-98.2\%) & 0.999 (0.996-1.000) & 0.923 (0.845-0.984) \\ \hline 
           Concat\_zero feature space& 95.4\% (89.7-100.0\%) & 0.999 (0.997-1.000) & 0.957 (0.899-1.000) \\ \hline \hline
           ImageNet-ResNet50 features& 83.7\% (73.2-92.9\%) & 
0.983 (0.965-0.996) & 0.849 (0.749-0.932) \\ \hline 
           
    \end{tabular}
    }
\end{table}

\begin{table}
    \centering
    \caption{Resulting p-values from paired two-tailed t-tests comparing each model to the baseline 5x + 10x graph model with UNI features and concat\_avg initialisation. These were calculated using the outputs of the five cross-validation models and were adjusted for multiple testing \cite{Benjamini1995}. Those less than 0.050 (before rounding) are indicated in \textbf{bold}.}
    \label{tab:pvals}
    \begin{tabular}{|c|c|c|c|c|} \hline 
         \makecell{\hspace{0.1cm} Validation \hspace{0.1cm} \\ p-values}&  Model&  \makecell{\hspace{0.1cm} Balanced \hspace{0.1cm} \\ Accuracy} &  \hspace{0.1cm} AUROC \hspace{0.1cm} & \hspace{0.1cm} F1 Score \hspace{0.1cm} \\ 
         \hline
         \multirow{6}{*}{\centering \makecell{Cross-\\Validation}}&  ABMIL 10x only& 0.078 & 0.264 & 0.105 \\  
         &  Graph 10x only& \textbf{0.046} & 0.756 & \textbf{0.035} \\  
         &  Graph 10x + 20x& 0.093 & 0.398 & 0.067 \\  
         &  Naive feature space& 0.226 & 0.360 & 0.158 \\  
         &  Concat\_zero feature space& \textbf{0.046} & 0.264 & \textbf{0.035} \\  
         &  \hspace{0.1cm} ImageNet-ResNet50 features \hspace{0.1cm} & \textbf{0.050} & \textbf{0.049} & \textbf{0.035} \\ 
         \hline
         \multirow{6}{*}{\centering \makecell{Hold-out \\ Testing}}&  ABMIL 10x only& \textbf{0.015} & \textbf{0.023} & \textbf{0.020} \\  
         &  Graph 10x only& \textbf{0.015} & \textbf{0.044} & \textbf{0.020} \\  
         &  Graph 10x + 20x& 0.106 & \textbf{0.023} & 0.113 \\  
         &  Naive feature space& 0.106 & \textbf{0.017} & 0.136 \\  
         &  Concat\_zero feature space& 0.106 & 0.119 & 0.107 \\  
         &  \hspace{0.1cm} ImageNet-ResNet50 features \hspace{0.1cm} & \textbf{0.021} & \textbf{0.008} & \textbf{0.032} \\ 
         \hline
         \multirow{6}{*}{\centering \makecell{External \\ Validation}}&  ABMIL 10x only& 0.926 & 0.312 & 0.681 \\  
         &  Graph 10x only& 0.418 & 0.464 & 0.166 \\  
         &  Graph 10x + 20x& 0.926 & 0.312 & 0.681 \\  
         &  Naive feature space& 0.418 & 0.364 & 0.166 \\  
         &  Concat\_zero feature space& 0.926 & 0.312 & 0.681 \\  
         &  \hspace{0.1cm} ImageNet-ResNet50 features \hspace{0.1cm} & 0.103 & 0.272 & 0.114 \\ 
         \hline
    \end{tabular}
    
\end{table}

\end{document}